\documentclass[aps,prl,preprint,groupedaddress,showpacs]{revtex4}
\usepackage{graphicx}  
\usepackage{dcolumn}
\usepackage{bm}
\usepackage{verbatim}

\begin{document}

\preprint{}

\title{Magnetism and Magnetic Isomers in Free Chromium Clusters}

\author{F. W. Payne, Wei Jiang, and L. A. Bloomfield}
\email{lab3e@Virginia.edu}
\affiliation{Department of Physics, University of Virginia, Charlottesville, 
Virginia 22904}

\date{\today}

\begin{abstract}
We have used the Stern-Gerlach deflection technique to study magnetism in chromium clusters of 20--133 atoms. Between 60~K and 100~K, we observe that these clusters have large magnetic moments and respond superparamagnetically to applied magnetic fields. Using superparamagnetic theory, we have determined the moment per atom for each cluster size and find that it often far exceeds the moment per atom present anywhere in the bulk antiferromagnetic lattice. Remarkably, our cluster beam contains two magnetically distinguishable forms of each cluster size with $\ge$ 34 atoms. We attribute this observation to structural isomers. 
\end{abstract}


\maketitle


Even before the rise of nanoscience, atomic clusters were of keen interest to people trying to understand how the physical properties of atoms and molecules evolve with size and shape into those of bulk materials. Clusters are true many-body quantum systems, too large for exact solutions in ordinary space and too small for exact solutions in momentum space, and the properties they exhibit are often rich and complex.

One such property, magnetism, offers unparalleled insight into clusters because it is sensitive to electronic and spatial structure, quantum size effects, surface to volume ratio, symmetry, and even temperature. Magnetism has been studied in low-dimensional supported systems (e.g., powders, granular metals, surfaces, and films) since Neel's pioneering work in the late 1940s and in isolated clusters since 1990.\cite{neel1949,bean1955,deheer1990}

Because clusters have reduced average coordination numbers relative to bulk and can adopt unusual shapes and symmetries, clusters of the ferromagnetic 3{\it d} transition metals were expected to have enhanced magnetism.\cite{liu1991,dunlap1991} Experimental studies of iron, cobalt, and nickel clusters do indeed find enhanced magnetism,\cite{bucher1991,billas1993, billas1994,apsel1996,xu2005} although their giant magnetic moments are partly obscured by the thermal fluctuation phenomenon known as superparamagnetism.\cite{neel1949,bean1955,khanna1991}.

Even more dramatic enhancements of magnetism were predicted in clusters of rhodium\cite{galicia1985,reddy1993} and manganese,\cite{pederson1998} 4{\it d} and 3{\it d} transition metals that are merely paramagnetic in the bulk. Experimental measurements of isolated rhodium and manganese clusters find large magnetic moments in small clusters of these two elements.\cite{cox1993,cox1994,knickelbein2001} 

Chromium, another of the 3{\it d} transition metals, is an itinerant antiferromagnet in the bulk. Below its $\sim$311~K N\'eel temperature, bulk chromium exhibits a transverse spin-density wave (SDW) that becomes a longitudinal spin-density wave below $\sim$123~K.\cite{fawcett1988} At zero temperature, chromium's SDW has a moment per atom of 0.43~$\mu_B$ rms and 0.62~$\mu_B$ peak. If chromium clusters had this same itinerant antiferromagnetic order, as though they were simply portions of the bulk {\it bcc} lattice, their moments per atom could not exceed 0.62~$\mu_B$.

Even 0.62~$\mu_B$ per atom is small compared to what is observed in iron, cobalt, and nickel clusters.\cite{bucher1991,douglass1993,billas1994,apsel1996} It came as no surprise then when an early study in our laboratory was unable to detect magnetism in chromium clusters of 9--31 atoms and set an upper limit of 0.77~$\mu_B$ per atom for their magnetic moments.\cite{douglass1992} More recently, chromium clusters supported on Au surfaces were shown to be antiferromagnetic.\cite{boeglin2005} Nonetheless, several theoretical studies predicted that small chromium clusters should be magnetic.\cite{salahub1981,pastor1989,lee1993,cheng1996,kohl1999,reddy1999}

In this paper, we report the results of a new Stern-Gerlach deflection study of chromium clusters in a cluster beam. Between 60~K and 100~K, all of the clusters we studied (20--133 atoms) have non-zero magnetic moments and deflect toward strong magnetic field. In many cases, the clusters are too magnetic to be simple fragments of the bulk lattice. Furthermore, our cluster beam contains no less than two magnetically distinguishable forms of each cluster size with 34 atoms or more. The most plausible explanation for these magnetically distinct populations is structural isomers---clusters with equal numbers of chromium atoms but different crystal structures.

The possibility that chromium clusters occur in electronically distinguishable isomers was suggested by Nickelbein, when he observed two distinct ionization potentials for Cr$_8$ and Cr$_{18}$.\cite{knickelbein2003}. Knickelbein did not report on any clusters larger than Cr$_{25}$, however, and an earlier photoelectron spectroscopy study by Wang reported no evidence for isomers between Cr$_1$ and Cr$_{55}$.\cite{wang1997}

In outline, our experiment consists of producing a collimated and temperature-controlled beam of chromium clusters, deflecting that beam with a gradient-field magnet, and determining the masses and deflections of the clusters downstream from the magnet. While the apparatus and techniques we used in the present study are similar to those used previously in our laboratory,\cite{bucher1993} a number of technical improvements have greatly enhanced the overall sensitivity and reproducibility of the measurements. In particular, a new cluster source offers more complete control of cluster temperatures and an automated data acquisition protocol allows us to take many different measurements concurrently and thereby reduce systematic errors due to small but unavoidable long-term drifts in the apparatus.

Chromium clusters form when focused 2nd-harmonic light (532~nm) from a Nd:YAG laser vaporizes part of a chromium sample rod in the presence of a pulse of dense helium gas. The resulting mixture of chromium vapor and helium then flows through a temperature-controlled tube, 2.5~mm in diameter and 150~mm long, before expanding supersonically into vacuum through a 1~mm diameter nozzle.

The long flow tube is critical to controlling cluster temperature.\cite{collings1993} Because the chromium/helium mixture spends $\sim$1~ms or more passing through this tube, there is time for cluster growth to die away and for the chromium clusters and helium gas to come into thermal equilibrium with the tube itself. The seeded supersonic expansion that subsequently produces the cluster beam is quite effective at cooling the clusters' translational temperatures and is moderately effective at cooling their rotational temperatures. But the expansion is too brief and too inefficient to have any significant effect on the clusters' vibrational temperatures. For all practical purposes, the chromium clusters' vibrational temperatures remain in equilibrium with the flow tube itself.

The chromium cluster beam is collimated by two narrow slits and a chopper wheel before entering the deflecting magnet. That electromagnet is a quadrant of a quadrupole,\cite{mccolm1966} 254-mm long, with a transverse field of up to 0.950~T and a transverse field gradient of up to 360~T/m. While in the magnet, each cluster experiences a force $\nabla(N{\bm \mu\cdot}{\bf B})$, where ${\bm\mu}$ is the cluster's magnetic moment per atom, $N$ is the number of atoms in the cluster, and ${\bf B}$ is the magnetic field. Since the force is perpendicular to the beam, the clusters deflect.

After the magnet, clusters pass through a field-free drift region, 1.183-m long, and into the ionization region of a time-of-flight mass spectrometer. There, a spatially filtered ArF excimer laser pulse (193~nm), directed antiparallel to the cluster beam, ionizes a narrow slice of that beam for mass analysis by the spectrometer. By scanning the laser light across the range of possible deflections and recording the cluster populations found at each location, we accumulate deflection profiles for many clusters sizes simultaneously. Small adjustments in timings and ion optics were necessary, however, to gather deflection profiles for the full range of cluster sizes reported in this paper (20--133 atoms).

Chromium clusters, like all transition metal clusters studied previously in beams, deflect solely toward strong field.\cite{bucher1991,billas1993,billas1994,apsel1996,xu2005} Such behavior is consistent with superparamagnetism, a thermal relaxation phenomenon in which a cluster's internal magnetic moment fluctuates rapidly in orientation and explores the entire Boltzmann distribution in a time short compared to the measurement time. For an isolated cluster, vibrational modes act as the heat bath responsible for thermal fluctuations and its vibrational temperature therefore appears in the Boltzmann factor. Since the Boltzmann distribution favors orientations that are aligned with an applied magnetic field, a superparamagnetic particle exhibits a time-averaged magnetic moment that is aligned with an applied magnetic field and that increases as that field increases. 

This time-averaged magnetic moment is what our deflection experiment measures. Assuming that the cluster's superparamagnetic moment is large enough to be treated classically and that all moment orientations are equal in energy in the absence of an applied field, the time-averaged or effective magnetic moment per atom ($\mu_{\rm eff}$) is given by the Langevin function:
\begin{eqnarray}
\mu_{\rm eff} & = & \mu{\cal L}(N\mu B/kT_{\rm vib})\label{eq:langevin}\\
& = & \mu[\coth(N\mu B/kT_{\rm vib})-(kT_{\rm vib}/N\mu B)]\nonumber
\end{eqnarray}
\noindent where $B$ is the magnetic field and $T_{\rm vib}$ is the vibrational temperature of the cluster, which we take to be the temperature of the source. Having determined a cluster's measured magnetic moment per atom $\mu_{\rm eff}$ from its deflection in a specific field at a specific vibrational temperature, we invert Eq.~(\ref{eq:langevin}) to obtain $\mu$, the magnitude of the cluster's true magnetic moment per atom, ${\bm\mu}$.

Because the magnetic moments of the chromium clusters were relatively small, we made all of the measurements at low temperatures (60~K to 100~K) and high fields (0.5~T to 1~T). Over this range of temperatures and fields, we saw no deviation from superparamagnetism in any of the chromium clusters---the measured moments per atom were always consistent with Eq.~({\ref{eq:langevin}).

We did observe, however, the separation of deflecting chromium clusters into two distinct populations, one significantly more magnetic than the other. Cr$_{41}$, shown in Fig.~\ref{fig:profiles}, is one of many cluster sizes exhibiting this behavior. When we increase the magnetic field and field-gradient, the spatial profile of Cr$_{41}$ clusters divides into two partially resolved peaks. Each peak separately obeys Eq.~(\ref{eq:langevin}), but with its own magnetic moment per atom, $\mu$. About 58\% of the Cr$_{41}$ clusters have $\mu=0.40\pm0.01~\mu_B$ and about 42\% have $\mu=1.00\pm0.01~\mu_B$. 

\begin{figure}
\includegraphics{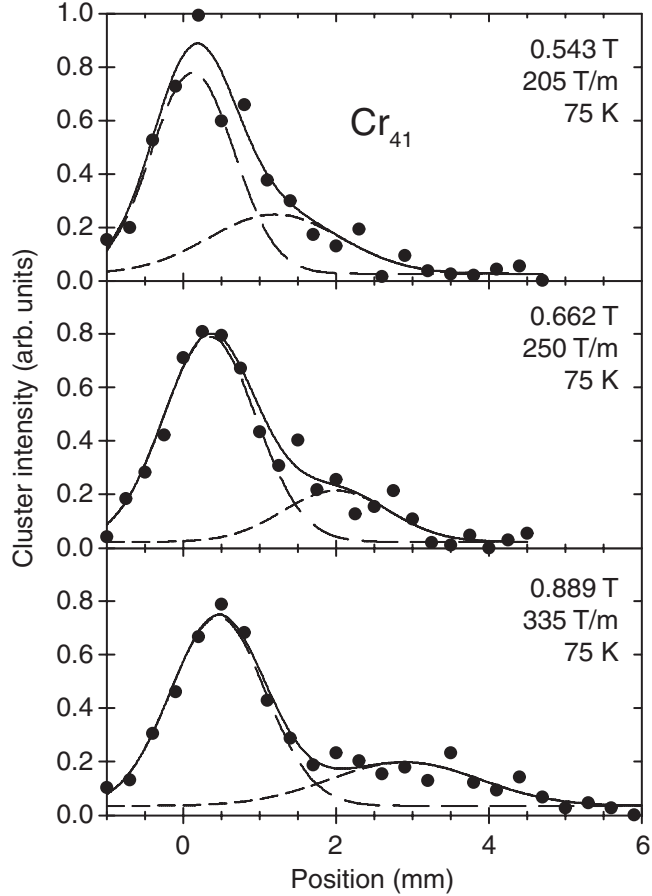}
\caption{\label{fig:profiles} Cr$_{41}$ clusters separate into two magnetically distinct populations after passing through a gradient magnetic field. Experimental data points are indicated. The solid curve is a fit to those points, constructed by summing the two Gaussians indicated by the dashed curves. While each population behaves superparamagnetically, they have different magnetic moments per atom, 0.40~$\mu_B$ and 1.00~$\mu_B$. As the field and field gradient increase, the two deflection peaks corresponding to those populations separate from one another.}
\end{figure}

We have never observed such magnetically distinguishable cluster populations in any of the other transition metals, despite considerable effort in that regard. A study of cobalt clusters performed concurrently with this chromium study observed only single deflection peaks for all cobalt clusters at all temperatures and all magnetic fields.\cite{payne2006} While superparamagnetic deflection peaks always broaden with increasing deflection, due primarily to the thermal statistics of finite systems,\cite{knickelbein2004} so far only chromium clusters have exhibited multiple superparamagnetic deflection peaks.

Figure~\ref{fig:moments} shows the magnetic moment per atom $\mu$ for chromium clusters of 20--133 atoms. For Cr$_{30}$ and Cr$_{34}$--Cr$_{133}$, we observe two peaks in the deflection profile and therefore report two values for $\mu$. The relative abundances of clusters in those two peaks are represented in Fig.~\ref{fig:moments} by the relative areas of their marker circles.

\begin{figure*}
\includegraphics{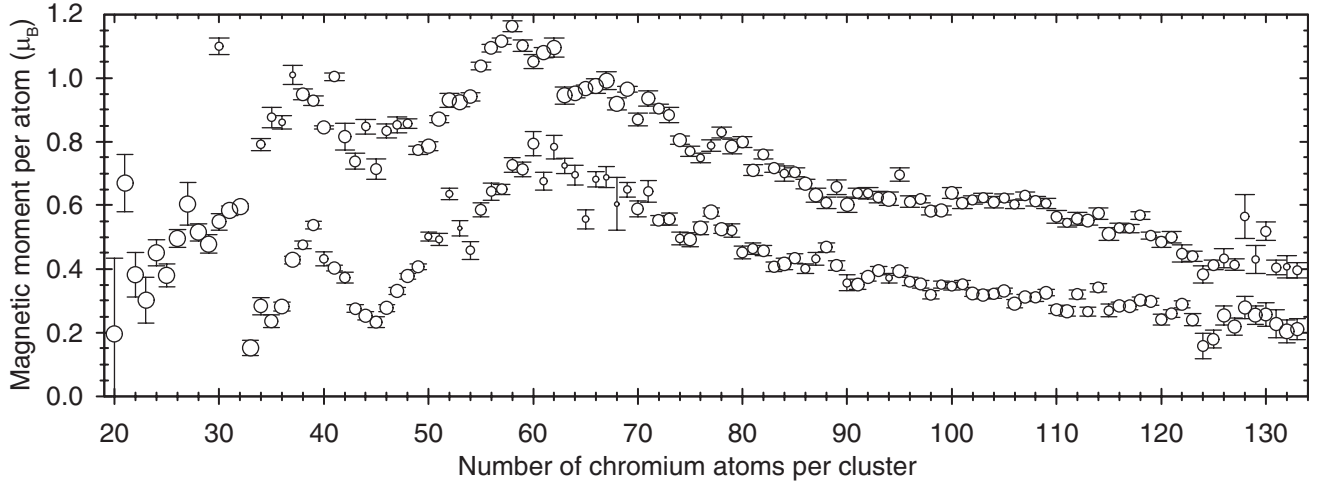}
\caption{\label{fig:moments} Magnetic moments per atom $\mu$ for chromium clusters of 20--133 atoms. Cr$_{30}$ and Cr$_{34}$--Cr$_{133}$ each has two magnetically distinguishable isomers so two value of $\mu$ are shown. The areas of the circles marking those two $\mu$ values are proportional to the populations of their respective isomers in the cluster beam. The error bars indicate statistical uncertainties (1 SD). Systematic uncertainties (not shown) are $\pm$0.1 $\mu_B$.}
\end{figure*}

The onset of magnetically distinguishable clusters at Cr$_{34}$ follows a dramatic drop in magnetism at Cr$_{33}$. Below that size, only Cr$_{30}$ shows unambiguous evidence of a separable population. Above that size, two sequences of chromium clusters are present in the beam, one significantly more magnetic than the other. The magnetic moments per atom and the relative populations of those two species evolve gradually across the size spectrum, with the high-moment species dominating Cr$_{50}$--Cr$_{71}$.

We have tried without success to explain these two populations as artifacts of the experiment. The two-peak deflection profiles are robust and reproducible, and also appeared in preliminary data taken with a very different type of cluster source.\cite{bloomfield2000} Extensive experimentation with source parameters has established that the two populations do not differ simply in temperature. The complicated and reproducible evolution of relative populations and the surprisingly large moments per atom of the high-moment species eliminate the possibility that one population consists of agglomerated smaller clusters. The mass resolution in the spectrometer is sufficient to determine that the clusters are pure chromium, free of oxygen atoms or other important contaminants. The possibility that one population involves a low-lying metastable electronic state that persists across a vast range of cluster size seems remote.

The most plausible explanation for these distinct populations is structural isomers---clusters that differ in their underlying crystalline structures. Trivial isomers, in which individual surface atoms are shifted about, are unlikely to cause such large differences in moment or to affect such a broad range of clusters so similarly. Therefore we suggest that the two population sequences are the result of two different crystal lattices. For example, the low-moment sequence might resemble the {\it bcc} lattice of bulk chromium and the high-moment sequence might be {\it fcc}-like, {\it hcp}-like, or even icosahedral.

In most of the low-moment chromium clusters, $\mu$ lies near or below the peak moment per atom for the SDW in bulk chromium (0.62 $\mu_B$). The wavelength of that SDW at 78~K is about 21 {\it bcc} unit cells, so the clusters are small enough to contain just one crest of the SDW. The gradual decrease in $\mu$ with increasing cluster size could then be explained as the inclusion of more of the SDW.

The high-moment chromium clusters, however, cannot be fragments of the bulk {\it bcc} chromium crystal. Cr$_{58}$ has a $\mu$ of 1.16$\pm$0.02~$\mu_B$, making it almost twice as magnetic as bulk nickel and more than 70\% as magnetic as bulk cobalt. This remarkably magnetic form of chromium exists for all clusters from Cr$_{34}$ up to at least Cr$_{133}$, and it is possible that it exists for smaller clusters as well. While preliminary measurements showed evidence for the high-moment cluster all the way down to Cr$_{9}$,\cite{bloomfield2000} the more careful measurements presented here were only able to confirm the existence of a high-moment species for Cr$_{30}$.

The periodic rises and falls in $\mu$ across each of the two sequences are most likely due to geometric shell structure. Studies in other transition metal clusters have observed minima in $\mu$ near shell closings, where the clusters are most compact and bulk-like, and maxima in $\mu$ between shells closings, where the clusters are least compact.\cite{xie2003,fujima1996,billas1994} That the maxima and minima in $\mu$ occur at different sizes in the two sequences may indicate that their shell closings are different.

In summary, we have observed magnetism in chromium clusters of 20--133 atoms at temperatures between 60~K and 100~K. For Cr$_{30}$, Cr$_{34}$-Cr$_{133}$, we find two magnetically distinguishable populations, with the high-moment species being far more magnetic on a per-atom basis than anything in bulk chromium. We propose that these two populations are the result of structural isomers, with the high-moment clusters having an underlying crystal structure different from that of bulk chromium.

\begin{acknowledgments}
This material is based upon work supported by the National Science Foundation under Grant No. DMR-0405203.
\end{acknowledgments}

\bibliography{manuscript}

\begin{thebibliography}{37}
\expandafter\ifx\csname natexlab\endcsname\relax\def\natexlab#1{#1}\fi
\expandafter\ifx\csname bibnamefont\endcsname\relax
  \def\bibnamefont#1{#1}\fi
\expandafter\ifx\csname bibfnamefont\endcsname\relax
  \def\bibfnamefont#1{#1}\fi
\expandafter\ifx\csname citenamefont\endcsname\relax
  \def\citenamefont#1{#1}\fi
\expandafter\ifx\csname url\endcsname\relax
  \def\url#1{\texttt{#1}}\fi
\expandafter\ifx\csname urlprefix\endcsname\relax\def\urlprefix{URL }\fi
\providecommand{\bibinfo}[2]{#2}
\providecommand{\eprint}[2][]{\url{#2}}

\bibitem[{\citenamefont{N\'eel}(1949)}]{neel1949}
\bibinfo{author}{\bibfnamefont{L.}~\bibnamefont{N\'eel}},
  \bibinfo{journal}{Compt. Rend.} \textbf{\bibinfo{volume}{228}},
  \bibinfo{pages}{664} (\bibinfo{year}{1949}).

\bibitem[{\citenamefont{Bean}(1955)}]{bean1955}
\bibinfo{author}{\bibfnamefont{C.~P.} \bibnamefont{Bean}}, \bibinfo{journal}{J.
  Appl. Phys.} \textbf{\bibinfo{volume}{26}}, \bibinfo{pages}{1381}
  (\bibinfo{year}{1955}).

\bibitem[{\citenamefont{de~Heer et~al.}(1990)\citenamefont{de~Heer, Milani, and
  Chatelain}}]{deheer1990}
\bibinfo{author}{\bibfnamefont{W.~A.} \bibnamefont{de~Heer}},
  \bibinfo{author}{\bibfnamefont{P.}~\bibnamefont{Milani}}, \bibnamefont{and}
  \bibinfo{author}{\bibfnamefont{A.}~\bibnamefont{Chatelain}},
  \bibinfo{journal}{Phys. Rev. Lett.} \textbf{\bibinfo{volume}{65}},
  \bibinfo{pages}{488} (\bibinfo{year}{1990}).

\bibitem[{\citenamefont{Liu et~al.}(1991)\citenamefont{Liu, Khanna, and
  Jena}}]{liu1991}
\bibinfo{author}{\bibfnamefont{F.}~\bibnamefont{Liu}},
  \bibinfo{author}{\bibfnamefont{S.~N.} \bibnamefont{Khanna}},
  \bibnamefont{and} \bibinfo{author}{\bibfnamefont{P.}~\bibnamefont{Jena}},
  \bibinfo{journal}{Phys.\ Rev.\ B} \textbf{\bibinfo{volume}{43}},
  \bibinfo{pages}{8179} (\bibinfo{year}{1991}).

\bibitem[{\citenamefont{Dunlap}(1991)}]{dunlap1991}
\bibinfo{author}{\bibfnamefont{B.~I.} \bibnamefont{Dunlap}},
  \bibinfo{journal}{Z.\ Phys.\ D} \textbf{\bibinfo{volume}{19}},
  \bibinfo{pages}{255} (\bibinfo{year}{1991}).

\bibitem[{\citenamefont{Bucher et~al.}(1991)\citenamefont{Bucher, Douglass, and
  Bloomfield}}]{bucher1991}
\bibinfo{author}{\bibfnamefont{J.~P.} \bibnamefont{Bucher}},
  \bibinfo{author}{\bibfnamefont{D.~C.} \bibnamefont{Douglass}},
  \bibnamefont{and} \bibinfo{author}{\bibfnamefont{L.~A.}
  \bibnamefont{Bloomfield}}, \bibinfo{journal}{Phys.\ Rev.\ Lett.}
  \textbf{\bibinfo{volume}{66}}, \bibinfo{pages}{3052} (\bibinfo{year}{1991}).

\bibitem[{\citenamefont{Billas et~al.}(1993)\citenamefont{Billas, Becker,
  Chatelain, and de\ Heer}}]{billas1993}
\bibinfo{author}{\bibfnamefont{I.~M.~L.} \bibnamefont{Billas}},
  \bibinfo{author}{\bibfnamefont{J.~A.} \bibnamefont{Becker}},
  \bibinfo{author}{\bibfnamefont{A.}~\bibnamefont{Chatelain}},
  \bibnamefont{and} \bibinfo{author}{\bibfnamefont{W.~A.} \bibnamefont{de\
  Heer}}, \bibinfo{journal}{Phys.\ Rev.\ Lett.} \textbf{\bibinfo{volume}{71}},
  \bibinfo{pages}{4067} (\bibinfo{year}{1993}).

\bibitem[{\citenamefont{Billas et~al.}(1994)\citenamefont{Billas, Chatelain,
  and de\ Heer}}]{billas1994}
\bibinfo{author}{\bibfnamefont{I.~M.~L.} \bibnamefont{Billas}},
  \bibinfo{author}{\bibfnamefont{A.}~\bibnamefont{Chatelain}},
  \bibnamefont{and} \bibinfo{author}{\bibfnamefont{W.~A.} \bibnamefont{de\
  Heer}}, \bibinfo{journal}{Science} \textbf{\bibinfo{volume}{265}},
  \bibinfo{pages}{1682} (\bibinfo{year}{1994}).

\bibitem[{\citenamefont{Apsel et~al.}(1996)\citenamefont{Apsel, Emmert, Deng,
  and Bloomfield}}]{apsel1996}
\bibinfo{author}{\bibfnamefont{S.~E.} \bibnamefont{Apsel}},
  \bibinfo{author}{\bibfnamefont{J.~W.} \bibnamefont{Emmert}},
  \bibinfo{author}{\bibfnamefont{J.}~\bibnamefont{Deng}}, \bibnamefont{and}
  \bibinfo{author}{\bibfnamefont{L.~A.} \bibnamefont{Bloomfield}},
  \bibinfo{journal}{Phys.\ Rev.\ Lett.} \textbf{\bibinfo{volume}{76}},
  \bibinfo{pages}{1441} (\bibinfo{year}{1996}).

\bibitem[{\citenamefont{Xu et~al.}(2005)\citenamefont{Xu, Yin, Moro, and de\
  Heer}}]{xu2005}
\bibinfo{author}{\bibfnamefont{X.}~\bibnamefont{Xu}},
  \bibinfo{author}{\bibfnamefont{S.}~\bibnamefont{Yin}},
  \bibinfo{author}{\bibfnamefont{R.}~\bibnamefont{Moro}}, \bibnamefont{and}
  \bibinfo{author}{\bibfnamefont{W.~A.} \bibnamefont{de\ Heer}},
  \bibinfo{journal}{Phys.\ Rev.\ Lett.} \textbf{\bibinfo{volume}{95}},
  \bibinfo{pages}{237209} (\bibinfo{year}{2005}).

\bibitem[{\citenamefont{Khanna and Linderoth}(1991)}]{khanna1991}
\bibinfo{author}{\bibfnamefont{S.~N.} \bibnamefont{Khanna}} \bibnamefont{and}
  \bibinfo{author}{\bibfnamefont{S.}~\bibnamefont{Linderoth}},
  \bibinfo{journal}{Phys. Rev. Lett.} \textbf{\bibinfo{volume}{67}},
  \bibinfo{pages}{742} (\bibinfo{year}{1991}).

\bibitem[{\citenamefont{Galacia}(1985)}]{galicia1985}
\bibinfo{author}{\bibfnamefont{R.}~\bibnamefont{Galacia}},
  \bibinfo{journal}{Rev. Mex. Fis.} \textbf{\bibinfo{volume}{32}},
  \bibinfo{pages}{51} (\bibinfo{year}{1985}).

\bibitem[{\citenamefont{Reddy et~al.}(1993)\citenamefont{Reddy, Khanna, and
  Dunlap}}]{reddy1993}
\bibinfo{author}{\bibfnamefont{B.~V.} \bibnamefont{Reddy}},
  \bibinfo{author}{\bibfnamefont{S.~N.} \bibnamefont{Khanna}},
  \bibnamefont{and} \bibinfo{author}{\bibfnamefont{B.~I.}
  \bibnamefont{Dunlap}}, \bibinfo{journal}{Phys. Rev. Lett.}
  \textbf{\bibinfo{volume}{70}}, \bibinfo{pages}{3323} (\bibinfo{year}{1993}).

\bibitem[{\citenamefont{Pederson et~al.}(1998)\citenamefont{Pederson, Reuse,
  and Khanna}}]{pederson1998}
\bibinfo{author}{\bibfnamefont{M.~R.} \bibnamefont{Pederson}},
  \bibinfo{author}{\bibfnamefont{F.}~\bibnamefont{Reuse}}, \bibnamefont{and}
  \bibinfo{author}{\bibfnamefont{S.~N.} \bibnamefont{Khanna}},
  \bibinfo{journal}{Phys. Rev. B} \textbf{\bibinfo{volume}{58}},
  \bibinfo{pages}{5632} (\bibinfo{year}{1998}).

\bibitem[{\citenamefont{Cox et~al.}(1993)\citenamefont{Cox, Louderback, and
  Bloomfield}}]{cox1993}
\bibinfo{author}{\bibfnamefont{A.~J.} \bibnamefont{Cox}},
  \bibinfo{author}{\bibfnamefont{J.~G.} \bibnamefont{Louderback}},
  \bibnamefont{and} \bibinfo{author}{\bibfnamefont{L.~A.}
  \bibnamefont{Bloomfield}}, \bibinfo{journal}{Phys. Rev. Lett.}
  \textbf{\bibinfo{volume}{71}}, \bibinfo{pages}{923} (\bibinfo{year}{1993}).

\bibitem[{\citenamefont{Cox et~al.}(1994)\citenamefont{Cox, Louderback, Apsel,
  and Bloomfield}}]{cox1994}
\bibinfo{author}{\bibfnamefont{A.~J.} \bibnamefont{Cox}},
  \bibinfo{author}{\bibfnamefont{J.~G.} \bibnamefont{Louderback}},
  \bibinfo{author}{\bibfnamefont{S.~E.} \bibnamefont{Apsel}}, \bibnamefont{and}
  \bibinfo{author}{\bibfnamefont{L.~A.} \bibnamefont{Bloomfield}},
  \bibinfo{journal}{Phys.\ Rev.\ B} \textbf{\bibinfo{volume}{49}},
  \bibinfo{pages}{12295} (\bibinfo{year}{1994}).

\bibitem[{\citenamefont{Knickelbein}(2001)}]{knickelbein2001}
\bibinfo{author}{\bibfnamefont{M.~B.} \bibnamefont{Knickelbein}},
  \bibinfo{journal}{Phys.\ Rev.\ Lett.} \textbf{\bibinfo{volume}{86}},
  \bibinfo{pages}{5255} (\bibinfo{year}{2001}).

\bibitem[{\citenamefont{Fawcett}(1988)}]{fawcett1988}
\bibinfo{author}{\bibfnamefont{E.}~\bibnamefont{Fawcett}},
  \bibinfo{journal}{Rev. Mod. Phys.} \textbf{\bibinfo{volume}{60}},
  \bibinfo{pages}{209} (\bibinfo{year}{1988}).

\bibitem[{\citenamefont{Douglass et~al.}(1993)\citenamefont{Douglass, Cox,
  Bucher, and Bloomfield}}]{douglass1993}
\bibinfo{author}{\bibfnamefont{D.~C.} \bibnamefont{Douglass}},
  \bibinfo{author}{\bibfnamefont{A.~J.} \bibnamefont{Cox}},
  \bibinfo{author}{\bibfnamefont{J.~P.} \bibnamefont{Bucher}},
  \bibnamefont{and} \bibinfo{author}{\bibfnamefont{L.~A.}
  \bibnamefont{Bloomfield}}, \bibinfo{journal}{Phys. Rev. B}
  \textbf{\bibinfo{volume}{47}}, \bibinfo{pages}{12874} (\bibinfo{year}{1993}).

\bibitem[{\citenamefont{Douglass et~al.}(1992)\citenamefont{Douglass, Bucher,
  and Bloomfield}}]{douglass1992}
\bibinfo{author}{\bibfnamefont{D.~C.} \bibnamefont{Douglass}},
  \bibinfo{author}{\bibfnamefont{J.~P.} \bibnamefont{Bucher}},
  \bibnamefont{and} \bibinfo{author}{\bibfnamefont{L.~A.}
  \bibnamefont{Bloomfield}}, \bibinfo{journal}{Phys.\ Rev.\ B}
  \textbf{\bibinfo{volume}{45}}, \bibinfo{pages}{6341} (\bibinfo{year}{1992}).

\bibitem[{\citenamefont{Boeglin et~al.}(2005)\citenamefont{Boeglin, Ohresser,
  Decker, Bulou, Scheurer, Chado, Dhesi, Gaudry, and Lazarovits}}]{boeglin2005}
\bibinfo{author}{\bibfnamefont{C.}~\bibnamefont{Boeglin}},
  \bibinfo{author}{\bibfnamefont{P.}~\bibnamefont{Ohresser}},
  \bibinfo{author}{\bibfnamefont{R.}~\bibnamefont{Decker}},
  \bibinfo{author}{\bibfnamefont{H.}~\bibnamefont{Bulou}},
  \bibinfo{author}{\bibfnamefont{F.}~\bibnamefont{Scheurer}},
  \bibinfo{author}{\bibfnamefont{I.}~\bibnamefont{Chado}},
  \bibinfo{author}{\bibfnamefont{S.~S.} \bibnamefont{Dhesi}},
  \bibinfo{author}{\bibfnamefont{E.}~\bibnamefont{Gaudry}}, \bibnamefont{and}
  \bibinfo{author}{\bibfnamefont{B.}~\bibnamefont{Lazarovits}},
  \bibinfo{journal}{Phys.\ Status\ Solidi B} \textbf{\bibinfo{volume}{242}},
  \bibinfo{pages}{1775} (\bibinfo{year}{2005}).

\bibitem[{\citenamefont{Salahub and Messmer}(1981)}]{salahub1981}
\bibinfo{author}{\bibfnamefont{D.~R.} \bibnamefont{Salahub}} \bibnamefont{and}
  \bibinfo{author}{\bibfnamefont{R.~P.} \bibnamefont{Messmer}},
  \bibinfo{journal}{Surf. Sci.} \textbf{\bibinfo{volume}{106}},
  \bibinfo{pages}{415} (\bibinfo{year}{1981}).

\bibitem[{\citenamefont{Pastor et~al.}(1989)\citenamefont{Pastor,
  Dorantes-Davila, and Bennemann}}]{pastor1989}
\bibinfo{author}{\bibfnamefont{G.~M.} \bibnamefont{Pastor}},
  \bibinfo{author}{\bibfnamefont{J.}~\bibnamefont{Dorantes-Davila}},
  \bibnamefont{and} \bibinfo{author}{\bibfnamefont{K.~H.}
  \bibnamefont{Bennemann}}, \bibinfo{journal}{Phys.\ Rev.\ B}
  \textbf{\bibinfo{volume}{40}}, \bibinfo{pages}{7642} (\bibinfo{year}{1989}).

\bibitem[{\citenamefont{Lee and Callaway}(1993)}]{lee1993}
\bibinfo{author}{\bibfnamefont{K.}~\bibnamefont{Lee}} \bibnamefont{and}
  \bibinfo{author}{\bibfnamefont{J.}~\bibnamefont{Callaway}},
  \bibinfo{journal}{Phys. Rev. B} \textbf{\bibinfo{volume}{48}},
  \bibinfo{pages}{15358} (\bibinfo{year}{1993}).

\bibitem[{\citenamefont{Cheng and Wang}(1996)}]{cheng1996}
\bibinfo{author}{\bibfnamefont{H.}~\bibnamefont{Cheng}} \bibnamefont{and}
  \bibinfo{author}{\bibfnamefont{L.-S.} \bibnamefont{Wang}},
  \bibinfo{journal}{Phys.\ Rev.\ Lett.} \textbf{\bibinfo{volume}{77}},
  \bibinfo{pages}{51} (\bibinfo{year}{1996}).

\bibitem[{\citenamefont{Kohl and Bertsch}(1999)}]{kohl1999}
\bibinfo{author}{\bibfnamefont{C.}~\bibnamefont{Kohl}} \bibnamefont{and}
  \bibinfo{author}{\bibfnamefont{G.~F.} \bibnamefont{Bertsch}},
  \bibinfo{journal}{Phys.\ Rev.\ B} \textbf{\bibinfo{volume}{60}},
  \bibinfo{pages}{4205} (\bibinfo{year}{1999}).

\bibitem[{\citenamefont{Reddy et~al.}(1999)\citenamefont{Reddy, Khanna, and
  Jena}}]{reddy1999}
\bibinfo{author}{\bibfnamefont{B.~V.} \bibnamefont{Reddy}},
  \bibinfo{author}{\bibfnamefont{S.~N.} \bibnamefont{Khanna}},
  \bibnamefont{and} \bibinfo{author}{\bibfnamefont{P.}~\bibnamefont{Jena}},
  \bibinfo{journal}{Phys. Rev. B} \textbf{\bibinfo{volume}{60}},
  \bibinfo{pages}{15597} (\bibinfo{year}{1999}).

\bibitem[{\citenamefont{Knickelbein}(2003)}]{knickelbein2003}
\bibinfo{author}{\bibfnamefont{M.~B.} \bibnamefont{Knickelbein}},
  \bibinfo{journal}{Phys.\ Rev.\ A} \textbf{\bibinfo{volume}{67}},
  \bibinfo{pages}{013202} (\bibinfo{year}{2003}).

\bibitem[{\citenamefont{Wang et~al.}(1997)\citenamefont{Wang, Wu, and
  Cheng}}]{wang1997}
\bibinfo{author}{\bibfnamefont{L.-S.} \bibnamefont{Wang}},
  \bibinfo{author}{\bibfnamefont{H.}~\bibnamefont{Wu}}, \bibnamefont{and}
  \bibinfo{author}{\bibfnamefont{H.}~\bibnamefont{Cheng}},
  \bibinfo{journal}{Phys.\ Rev.\ B} \textbf{\bibinfo{volume}{55}},
  \bibinfo{pages}{12884} (\bibinfo{year}{1997}).

\bibitem[{\citenamefont{Bucher and Bloomfield}(1993)}]{bucher1993}
\bibinfo{author}{\bibfnamefont{J.~P.} \bibnamefont{Bucher}} \bibnamefont{and}
  \bibinfo{author}{\bibfnamefont{L.~A.} \bibnamefont{Bloomfield}},
  \bibinfo{journal}{Int.\ J.\ Mod.\ Phys.\ B} \textbf{\bibinfo{volume}{7}},
  \bibinfo{pages}{1079} (\bibinfo{year}{1993}).

\bibitem[{\citenamefont{Collings et~al.}(1993)\citenamefont{Collings, Amrein,
  Rayner, and Hackett}}]{collings1993}
\bibinfo{author}{\bibfnamefont{B.~A.} \bibnamefont{Collings}},
  \bibinfo{author}{\bibfnamefont{A.~H.} \bibnamefont{Amrein}},
  \bibinfo{author}{\bibfnamefont{D.~M.} \bibnamefont{Rayner}},
  \bibnamefont{and} \bibinfo{author}{\bibfnamefont{P.~A.}
  \bibnamefont{Hackett}}, \bibinfo{journal}{J. Chem. Phys.}
  \textbf{\bibinfo{volume}{99}}, \bibinfo{pages}{4174} (\bibinfo{year}{1993}).

\bibitem[{\citenamefont{McColm}(1966)}]{mccolm1966}
\bibinfo{author}{\bibfnamefont{D.}~\bibnamefont{McColm}},
  \bibinfo{journal}{Rev.\ Sci.\ Instrum.} \textbf{\bibinfo{volume}{37}},
  \bibinfo{pages}{1115} (\bibinfo{year}{1966}).

\bibitem[{\citenamefont{Payne et~al.}(2006)\citenamefont{Payne, Jiang, Emmert,
  and Bloomfield}}]{payne2006}
\bibinfo{author}{\bibfnamefont{F.~W.} \bibnamefont{Payne}},
  \bibinfo{author}{\bibfnamefont{W.}~\bibnamefont{Jiang}},
  \bibinfo{author}{\bibfnamefont{J.~W.} \bibnamefont{Emmert}},
  \bibnamefont{and} \bibinfo{author}{\bibfnamefont{L.~A.}
  \bibnamefont{Bloomfield}} (\bibinfo{year}{2006}),
  \bibinfo{note}{unpublished}.

\bibitem[{\citenamefont{Knickelbein}(2004)}]{knickelbein2004}
\bibinfo{author}{\bibfnamefont{M.~B.} \bibnamefont{Knickelbein}},
  \bibinfo{journal}{J.\ Chem.\ Phys.} \textbf{\bibinfo{volume}{121}},
  \bibinfo{pages}{5281} (\bibinfo{year}{2004}).

\bibitem[{\citenamefont{Bloomfield et~al.}(2000)\citenamefont{Bloomfield, Deng,
  Zhang, and Emmert}}]{bloomfield2000}
\bibinfo{author}{\bibfnamefont{L.~A.} \bibnamefont{Bloomfield}},
  \bibinfo{author}{\bibfnamefont{J.}~\bibnamefont{Deng}},
  \bibinfo{author}{\bibfnamefont{H.}~\bibnamefont{Zhang}}, \bibnamefont{and}
  \bibinfo{author}{\bibfnamefont{J.~W.} \bibnamefont{Emmert}}, in
  \emph{\bibinfo{booktitle}{Proceedings of the International Symposium on
  Cluster and Nanostructure Interfaces}}, edited by
  \bibinfo{editor}{\bibfnamefont{P.}~\bibnamefont{Jena}},
  \bibinfo{editor}{\bibfnamefont{S.~N.} \bibnamefont{Khanna}},
  \bibnamefont{and} \bibinfo{editor}{\bibfnamefont{B.~K.} \bibnamefont{Rao}}
  (\bibinfo{publisher}{World, Singapore}, \bibinfo{year}{2000}), pp.
  \bibinfo{pages}{131--138}.

\bibitem[{\citenamefont{Xie and Blackman}(2003)}]{xie2003}
\bibinfo{author}{\bibfnamefont{Y.}~\bibnamefont{Xie}} \bibnamefont{and}
  \bibinfo{author}{\bibfnamefont{J.~A.} \bibnamefont{Blackman}},
  \bibinfo{journal}{J.\ Phys.: Condens.\ Matter} \textbf{\bibinfo{volume}{15}},
  \bibinfo{pages}{L615} (\bibinfo{year}{2003}).

\bibitem[{\citenamefont{Fujima and Yamaguchi}(1996)}]{fujima1996}
\bibinfo{author}{\bibfnamefont{N.}~\bibnamefont{Fujima}} \bibnamefont{and}
  \bibinfo{author}{\bibfnamefont{T.}~\bibnamefont{Yamaguchi}},
  \bibinfo{journal}{Phys.\ Rev.\ B} \textbf{\bibinfo{volume}{54}},
  \bibinfo{pages}{26} (\bibinfo{year}{1996}).

\end{thebibliography}

\end{document}